# Evidence for an excitonic insulator phase in a zero-gap InAs/GaSb bilayer

W. Yu[1], V. Clericò[2], C. Hernández Fuentevilla[2], X. Shi[1,§], Y. Jiang[3], D. Saha[4], W.K. Lou[5], K. Chang[5], D.H. Huang[6], G. Gumbs[7], D. Smirnov[8], C. J. Stanton[4], Z. Jiang[3], V. Bellani[9], Y. Meziani[2], E. Diez[2,*], W. Pan[1,*], S.D. Hawkins[1], J.F. Klem[1]

[1]*Sandia National Laboratories, Albuquerque, New Mexico 87185, USA*
[2]*Nanotechnology Group, Universidad de Salamanca, E-37008 Salamanca, Spain*
[3]*School of Physics, Georgia Institute of Technology, Atlanta, Georgia 30332, USA*
[4]*Department of Physics, University of Florida, Gainesville, FL 32611, USA*
[5]*SKLSM, Institute of Semiconductors, Chinese Academy of Sciences, 100083 Beijing, China*
[6]*Air Force Research Laboratory, Space Vehicles Directorate, Kirtland Air Force Base, NM 87117, USA*
[7]*Department of Physics and Astronomy, Hunter College of the City University of New York, 695 Park Avenue, New York, NY 10065, USA*
[8]*National High Magnetic Field Laboratory, Tallahassee, FL 32310, USA*
[9]*Dipartimento di Fisica, Università degli studi di Pavia, I-27100 Pavia, Italy*

Many-body interactions can produce novel ground states in a condensed-matter system. For example, interacting electrons and holes can spontaneously form excitons, a neutral bound state, provided that the exciton binding energy exceeds the energy separation between the single particle states. Here we report on electrical transport measurements on spatially separated two-dimensional electron and hole gases with nominally degenerate energy subbands, realized in an InAs(10 nm)/GaSb(5 nm) coupled quantum well. We observe a narrow and intense maximum (~500 kΩ) in the four-terminal resistivity in the charge neutrality region, separating the electron-like and hole-like regimes, with a strong activated temperature-dependence above $T$ = 7 K and perfect stability against quantizing magnetic fields. By quantitatively comparing our data with early theoretical predictions, we show that such unexpectedly large resistance in our nominally zero-gap semi-metal system is probably due to the formation of an excitonic insulator state.

More than 50 years ago, Mott made a seminal observation on the anomaly at the transition from a semi-metal to a semiconductor[1]. Later on, Knox and co-workers[2] developed this idea and argued that, if in a semiconductor (which at low temperature is insulating) the binding energy of the excitons ($E_B$) exceeds the energy gap ($E_g$), the conventional insulating ground state would be unstable against a new phase, dubbed excitonic insulator, which originates from the formation of an exciton condensate. The low temperature behavior of the conductivity for an excitonic insulator was theoretically predicted by Jerome and co-workers[3] and many other studies, both theoretical and experimental, have been performed afterwards[4-18]. Although experimental evidence on the existence of this unusual insulating state has been reported[7], no completely conclusive results have been obtained so far.

In recent years, new electronic materials, topological insulators (TIs), have been attracting increasing interest thanks to the possibility of hosting new exotic states of matter[19,20]. Among many TIs, InAs/GaSb double quantum wells (DQWs) are of particular interest, as they can readily be integrated within current semiconductor processing technologies. In this material system, the top of the valence band of GaSb is 0.143 eV higher than the bottom of the conduction band of InAs[21]. Due to the quantum confinement effect, the alignment of $E_0$ (the lowest electron subband in the conduction band) and $H_0$ (the highest hole subband in the valence band) can be tuned by varying the wells' thickness (an alternative way is represented by application of an electrical field[22]). For a given thickness of the GaSb QW, the bilayer is a conventional semiconductor provided that the thickness of the InAs QW is sufficiently small to keep $E_0$ above $H_0$[23]. Above a critical thickness of the InAs QW, a quantum spin Hall (QSH) phase is realized[24] due to the inversion between the electron and the hole ground levels. However, little is known about the transport properties at the critical thickness, which corresponds to the phase boundary between the normal insulator (NI) and the QSH insulator.

In this paper, we report on electrical transport experiments performed in InAs/GaSb DQWs in all three, normal, critical and inverted regimes. An unexpectedly huge resistance peak was observed in the charge neutrality region (CNR) in a sample where the $E_0$ and $H_0$ subbands almost perfectly touch each other at $k = 0$. We demonstrate that this huge resistance is due to the formation of an excitonic insulator phase at the charge neutrality point (CNP) in our critical-thickness device.

A series of InAs/GaSb DQW samples, in which the thickness of InAs ($d$) was varied from 8 nm to 15 nm while the thickness of GaSb was fixed at 5 nm, were grown with the molecular beam epitaxy (MBE) technique. An 8-band **k·p** method was employed to calculate the band structures of these DQW samples, using the Hamiltonian and basis functions of Ref. [35] and considering the strain effect resulting from the lattice mismatch between the GaSb substrate and each individual QW layer[36].

In Figure 1a we plot the evolution of the calculated band structures of the InAs/GaSb DQW at three different $d$ values, corresponding to the normal, critical and inverted regimes. We found that for $d < 10$ nm, $E_g > 0$, signaling a normal semiconductor band structure, while for $d > 10$ nm, $E_g < 0$, indicating the band-inverted regime of InAs/GaSb. At $d = 10$ nm $= d_c$ (critical thickness) we obtain $E_g = 0$, which corresponds to the boundary between the normal and inverted band structures. In the Supplementary Information we have reported two different methods to distinguish samples from normal, critical or inverted regimes confirming the 8-band **k·p** results showed in Figure 1. Recently, cyclotron resonance results were also reported in these three regimes[25].

In Figure 1b-d, we show the four-terminal resistance measured as a function of gate voltage ($V_g$) in three typical samples, where the InAs QW thickness is 9 nm (sample A), 10 nm (sample B), and 13 nm (sample C), respectively. The CNP peak was centered in each sample at $V_g = -0.60$ V (b), $V_g = -0.53$ V (c), and $V_g = -0.70$ V (d), respectively. Accordingly, the gate voltage is normalized so that the CNP corresponds to $V_g^* = 0$. In all three samples a resistance peak, centered at the CNP, separates two highly conductive regions associated to positively ($V_g^* < 0$ V) and negatively-charged carries ($V_g^* > 0$ V). The sign of the carriers in the two regimes was confirmed by Hall measurements at a finite magnetic field. Most strikingly, for $d = d_c = 10$ nm, where the electron and hole subbands are degenerate (middle panel of Figure 1a) and a semi-metal phase is expected, the 4-terminal resistance reaches an overwhelmingly large value ~530 kΩ in the CNR (Figure 1c, sample B). Measurements at both constant current and constant voltage bias measurements were carried out to confirm the value of this resistance peak. In contrast, in $d = 9$ and 13 nm DQWs, where the band gap is either positive or negative, the resistance in the CNR assumes smaller values: in the normal regime (Figure 1b, sample A), the resistance at the CNP is ~400 kΩ, while in the inverted regime (Figure 1d, sample C), the resistance is ~60 kΩ.

In the following, we will concentrate on the transport properties of this anomalous insulating state detected in sample B and will show that such high resistance at the critical regime is probably due to the formation of the excitonic insulator phase.

First, we observed that the maximum value of $R_{xx}$ shows very weak temperature dependence at low $T$ (see Fig. 2). This low temperature saturation behavior is different from the insulating state at the CNP in the single layer graphene[26]. The high-$T$ regime ($T > 7$ K) is properly described by an activated behavior $R_{xx}^{CNP} \propto e^{\Delta/2k_BT}$. The best fit of $R_{xx}$, shown in Figure 2b, gives an estimated energy gap $\Delta = 2.08 \pm 0.10$ meV. We note here this strongly activated $T$-dependence is in contrast with the reported logarithmic temperature dependence of inverted semiconductors[27,28]; it is also quite different from that found in normal semiconducting InAs/GaSb bilayers[23], where an energy gap of ~0.5 meV was reported, which is five times smaller than that measured in our sample. On the other hand, the value of the $R_{xx}$ peak at low temperature reported in Ref. [23] is one order of magnitude larger than ours. This clear discrepancy with respect to the NI case (strong *T*-dependence and lower $R_{xx}$) strongly suggests the realization of a different insulating phase in our degenerate sample.

Magnetotransport was further carried out in sample B. In Figure 3a we plot the longitudinal ($\sigma_{xx}$) and Hall (($\sigma_{xy}$) conductivities, measured at $B = 7$ T and $T = 100$ mK. Quantized plateaus at filling factors $\nu = 1$ and $\nu = 2$ (as well as at all higher values in the standard integer Quantum Hall sequence, not shown) are clearly developed for both electrons and holes. In the vicinity of the CNR, an additional plateau at $\sigma_{xy} = 0$ is observed; a similar quantized plateau was reported for an inverted InAs/GaSb sample in Ref. [29], as well as for a degenerate HgTe quantum well in Ref. [30].

In Figure 3b, we show the two-dimensional density of charge carriers for electrons *n* and holes *p* as a function of $V_g$. The experimental points (red and black open circles) were obtained from the position of the integer quantum Hall plateaus at several different magnetic fields, while the continuous lines are linear best fits. The lines of the fits cross at $V_g = -0.55$ V and at $n_0 = p_0 = -1.1 \times 10^{10}$ cm$^{-2}$ (which corresponds to the CNP condition $n + p = 0$). We note here that the obtained electron and hole densities at the CNP are the lowest reported, indicating a high quality of our sample. Figure 3c shows the normalized magnetoresistance ($R_{xx}(B) - R_{xx}(0))/R_{xx}(0)$ in the CNR at $T = 100$ mK: zero magnetoresistance is observed up to $B = 7$ T. Such stability under quantizing magnetic field extends also to its temperature-dependence, as shown in the inset of Figure 2b. Indeed we observed small differences between the data collected in the range 0 T< B < 7 T. These observations strongly differ from those reported for the insulating state at the CNP in Ref. [29], where a huge magnetoresistance accompanied by strong *B*-induced strengthening of the *T*-dependence was reported. In the Supplementary Information we have addressed this point carefully, by comparing the results for the critical sample with the normal and inverted ones where a larger dependence is observed.

On the basis of these observations, we conclude that the large resistance in the CNR in our critical sample is different from that observed in a NI in Ref. [23]. Rather, it is due to the formation of an excitonic insulator phase. When the Fermi level is tuned to the CNP, excitons will form by the attractive coulomb interactions between residual electrons and holes, whose densities are vanishingly small. Moreover, at the critical thickness $d_c$, the energy gap is expected to be almost zero and, thus, smaller than the exciton binding energy $E_B$, realizing the condition $E_g < E_B$. As predicted more than 50 years ago[1], the excitons under this condition condense, giving rise to an energy gap, which is responsible for the large resistance peak in the CNR. Away from the critical thickness, for $d < 10$ nm, a normal gap $E_g > E_B$ appears (see Figure S1(b) in the Supplementary Information). For $d > 10$ nm, the energy band structure becomes inverted. Due to strong interactions between electrons and holes, a mini gap ($\Delta$) is formed at a finite *k*. At the band edges, the two types of carries coexist. As a result, $E_B$ is much reduced[14]. Consequently, $E_B < \Delta$, preventing the formation of exciton condensation.

A comparison of these results with earlier calculations by Zhu *et al.*[31], further corroborates the formation of the excitonic insulating phase in our critical sample. Using the values quoted in Ref.[31], i.e., effective mass of $m_e = 0.023\ m_0$ and $m_h = 0.33\ m_0$ for electrons in InAs and holes in GaSb, and effective dielectric constant $\kappa \sim 15$, the calculated Bohr radius for exciton is $a_u \sim 36.9$ nm. As a result, $d/a_u = 0.271$, where $d$ is the InAs thickness. At this ratio, $E_B$ is much larger than $E_g$, which is close to 0 due to reduced screening, and the DQW is in the excitonic phase (see Figure 3 in Ref. [31]). Moreover, the exciton binding energy was calculated to be 2.13 meV, which is in excellent accordance with the energy gap $\Delta$ measured in our critical sample in the CNR. In the Supplementary Information (see Figure S4) we have obtained also a similar value for the binding energy of the excitonic phase from the perpendicular magnetic field dependence in our critical sample.

As a remark, we would like to mention a recent theoretical study of the charge transport in two-dimensional disordered semimetals by Knap *et al.*[32]. In that work it was found that electron and hole puddles, due to smooth fluctuations of the potential, are responsible for a large resistance peak in the CNR measured in HgTe QWs by Olshanetsky *et al.*[33]. First, we point out that the HgTe QW structure considered in Ref. [33], 20 nm wide, is far from the critical thickness of 6.3 nm in the HgTe QW[33]. As a result, the conduction and valence bands overlap is much greater than the gap estimated in our critical sample. Second, in Ref. [33] the authors found that the resistance at the CNP increases monotonically with decreasing temperature (Figure 3 in Ref. [33]), a behavior different from what we have observed. Finally, $R_{xx}$ shows relatively strong magnetic field dependence in Ref. [33] at the temperature considered, which, again, is very different from our observation (see Figure 3c).

Finally, while preparing this manuscript, we became aware of another work on gate-tuned spontaneous exciton insulator in InAs/GaSb double-quantum wells by Du *et al.*[34]. They also report on the observation of an excitonic insulating phase, by gating the residual electron and hole density at the CNP towards the diluted regime, which is naturally realized in our nominally zero-gap system. Notably, the activation gap as measured in their most diluted regime matches our value of $\Delta$.

This work was supported by the following projects: the U.S. Department of Energy, Office of Science, Basic Energy Sciences, Materials Sciences and Engineering Division at Sandia, MINECO and FEDER MAT2013-46308-C2-1-R and MAT2016-75955-C2-2-R, Junta de Castilla y León SA226U13 and SA045U16 at University of Salamanca. Y.J., D.S., and Z.J. acknowledge the support from the U.S. Department of Energy (Grant No. DE-FG02-07ER46451). Device fabrication was performed at the Center for Integrated Nanotechnologies, a

§Present address: Department of Physics, The University of Texas at Dallas, Richardson, TX 75080, USA.

Correspondence and requests for materials should be addressed to E.D. (email: enrisa@usal.es) and to W.P. (email:wpan@sandia.gov).

34. Du, L. Lou, W. Chang, K. Sullivan, G. & Du, R.-R. Gate-tuned spontaneous exciton insulator in double-quantum wells. *arXiv*:1508.04509 (2015).
35. Li, J., Yang, W. & Chang, K. Spin states in InAs/AlSb/GaSb semiconductor quantum wells. *Phys. Rev. B* **80**, 035303 (2009).
36. Novik, E. G. *et al.* Band structure of semimagnetic $Hg_{1-y}Mn_yTe$ quantum wells. *Phys. Rev. B* **72**, 035321 (2005).

# Figures and Figure Legends

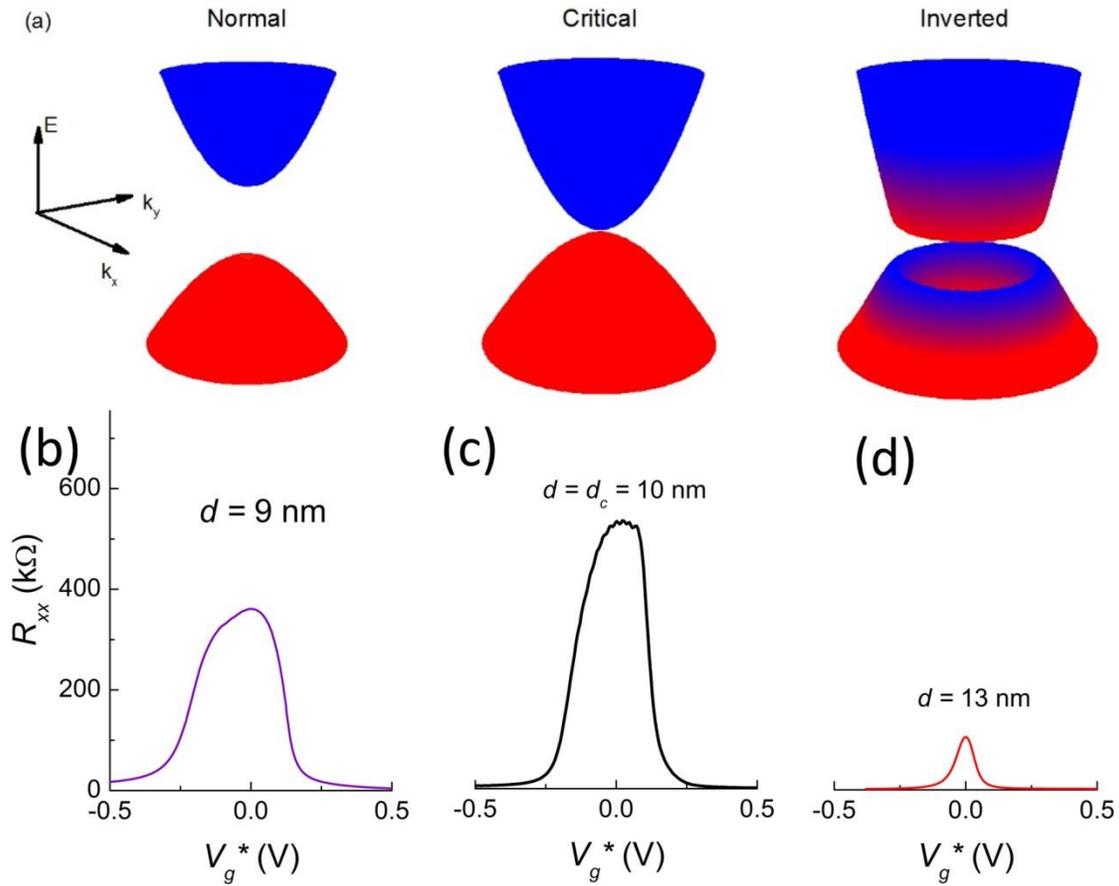

**Figure 1** (a) Band structures of the InAs/GaSb DQWs calculated using the 8-band **k·p** method (Supplementary Information) for three typical configurations: (left, sample A) $d$ = 9 nm, (middle, sample B) $d$ = 10 nm, and (right, sample C) $d$ =13 nm. (b-d) 4-terminal resistance as function of gate voltage, measured at $T$ = 500 mK, in samples A-C, respectively. The gate voltage is normalized so that the gate voltage at the CNP is zero ($V_g^* = V_g - V_g^{CNP}$).

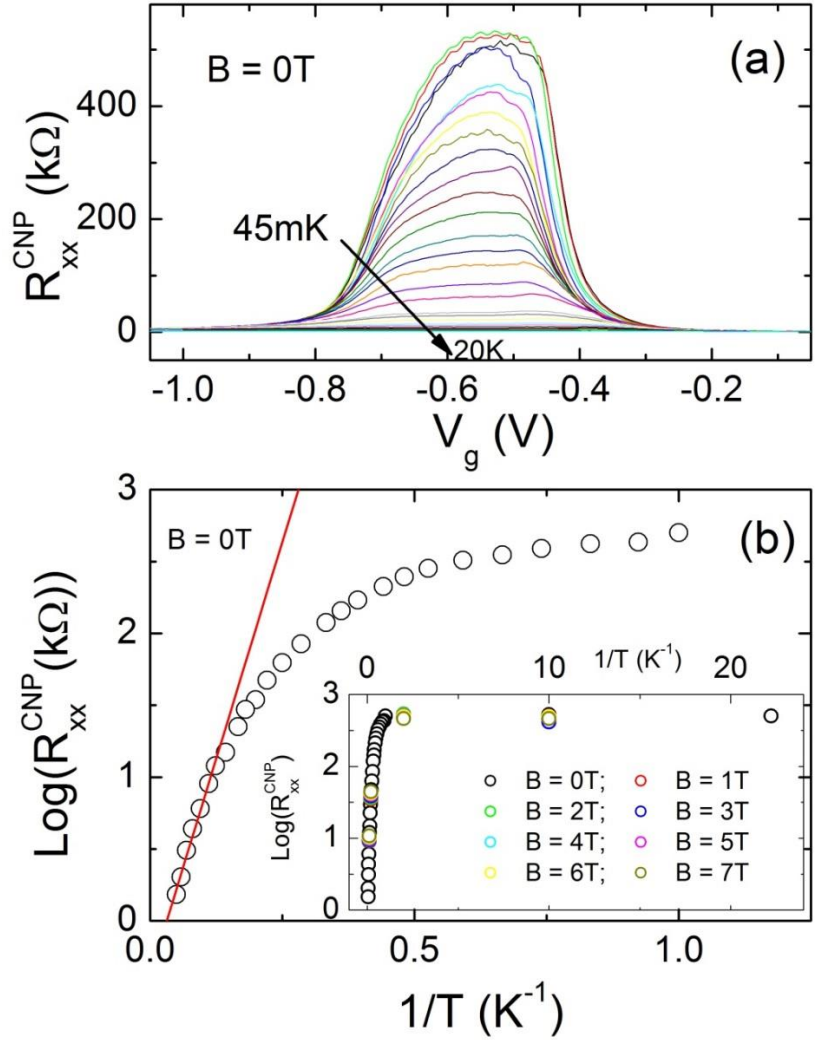

**Figure 2** (a) Longitudinal resistance $R_{xx}$ as a function of gate voltage $V_g$ for increasing temperatures at $B = 0$ T, measured in sample B. (b) maximum of $R_{xx}$ in the CNR as a function of $1/T$. The red line is a fit to $R_{xx}^{CNP} \propto e^{\Delta/2k_BT}$, obtained for the data at $T > 7$ K, which provides the estimation of the energy gap $\Delta$ reported in the text. The inset shows the temperature dependent data of $R_{xx}^{CNP}$ at various magnetic fields.

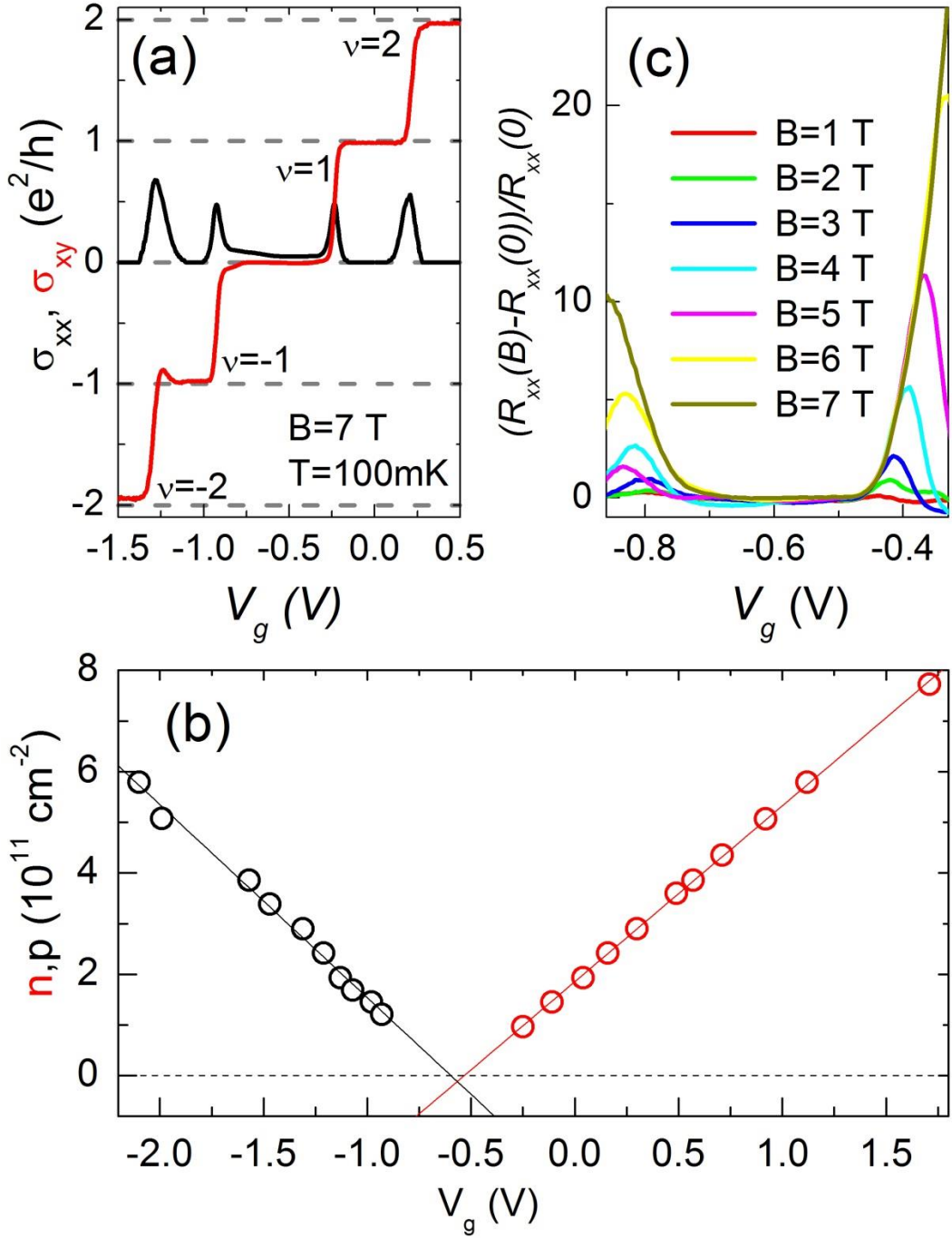

**Figure 3** (a) Longitudinal ($\sigma_{xx}$) and Hall ($\sigma_{xy}$) conductivities as a function of $V_g$, measured at $B = 7$ T. (b) Two-dimensional charge density for electrons $n$ and holes $p$ as a function of $V_g$. (c) Normalized magnetoresistance ($R_{xx}(B) - R_{xx}(0))/R_{xx}(0)$ in the CNR, for increasing magnetic fields. All the data were acquired at $T = 100$ mK.